\documentclass[conference]{IEEEtran}
\usepackage{listings}
\usepackage{xcolor} 

\definecolor{highlightcolor}{RGB}{255,255,204} 

\usepackage{graphicx}
\usepackage{booktabs}
\usepackage{multirow}
\usepackage{tabularx}
\usepackage{diagbox}
\usepackage{colortbl}
\usepackage{soul}
\usepackage{wrapfig}
\usepackage[caption=false]{subfig}
\usepackage[font=footnotesize,labelfont=bf]{caption}
\usepackage{dblfloatfix}
\usepackage{threeparttable} 

\usepackage{flushend}

\begin{document}

\title{Self-HWDebug: Automation of LLM Self-Instructing for Hardware Security Verification}

\author{\IEEEauthorblockN{Mohammad Akyash}
\IEEEauthorblockA{\textit{Dept. of Electrical and Comp. Eng. (ECE)} \\
\textit{University of Central Florida}\\
Orlando, US 32816 \\
mohammad.akyash@ucf.edu}
\and
\IEEEauthorblockN{Hadi Mardani Kamali}
\IEEEauthorblockA{\textit{Dept. of Electrical and Comp. Eng. (ECE)} \\
\textit{University of Central Florida}\\
Orlando, US 32816 \\
hadi.mardanikamali@ucf.edu}
}

\maketitle

\begin{abstract}

The rise of instruction-tuned Large Language Models (LLMs) marks a significant advancement in artificial intelligence (AI) (tailored to respond to specific prompts). Despite their popularity, applying such models to debug security vulnerabilities in hardware designs, i.e., register transfer language (RTL) modules, particularly at system-on-chip (SoC) level, presents considerable challenges. One of the main issues lies in the need for precisely designed instructions for pinpointing and mitigating the vulnerabilities, which requires substantial time and expertise from human experts. In response to this challenge, this paper proposes Self-HWDebug, an innovative framework that leverages LLMs to automatically create required debugging instructions. In Self-HWDebug, a set of already identified bugs from the most critical hardware common weakness enumeration (CWE) listings, along with mitigation resolutions, is provided to the framework, followed by prompting the LLMs to generate targeted instructions for such mitigation. The LLM-generated instructions are subsequently used as references to address vulnerabilities within the same CWE category but in totally different designs, effectively demonstrating the framework's ability to extend solutions across related security issues. Self-HWDebug significantly reduces human intervention by using the model's own output to guide debugging. Through comprehensive testing, Self-HWDebug proves not only to reduce experts' effort/time but also to even improve the quality of the debugging process.


\end{abstract}
\IEEEpeerreviewmaketitle

\section{Introduction}

Given the widespread use of SoCs within today's digital systems, coupled with the escalating size and complexity of their associated hardware, the emergence of unknown vulnerabilities (especially security vulnerabilities) stemming from their hardware has become an inevitable and challenging aspect of the integrated circuit (IC) supply chain process \cite{ray2017system}. To minimize re-spins due to post-silicon verification issues, addressing these vulnerabilities must be done at the highest level of abstraction, i.e., RTL \cite{ferraiuolo2017verification, hu2020overview}. This process is time-consuming and demands extensive hardware engineering expertise. Numerous methodologies have been investigated over the years for such challenges, from formal methods \cite{grimm2018survey} to advanced testing techniques, e.g., fuzzing \cite{azar2022fuzz, hossain2023socfuzzer}.  

More recently, significant advancements in seedling AI use, particularly through LLMs, have greatly enhanced the resolution of verification issues, particularly w.r.t. the automating the verification process, reducing required experts' knowledge and time \cite{akyash2024evolutionary}. Prompt engineering, which is the process of crafting inputs that guide LLMs' responses, has been widely used for hardware security verification purposes \cite{ahmad2024hardware, nair2023generating, orenes2023using}. For example, \cite{ahmad2024hardware} has explored employing expert-crafted prompts to steer model behavior in specific debugging scenarios. However, this method suffers from \textit{scalability} and \textit{prompts' efficacy} issues \cite{akyash2024evolutionary}. As hardware designs grow in complexity and the number of IP cores increases, the task of manually creating prompts that address every potential security vulnerability becomes impractical.  

One of the main shortcomings of manual prompt engineering for hardware security verification resides in the reliance upon prompts created by experts possessing deep expertise in a specific design or scenario. This approach often fails to consistently yield effective instructions tailored to address vulnerabilities. As a result, an instruction deemed effective for mitigating a given vulnerability within one design may fail when applied to the same vulnerability in a different design. This inherent limitation adversely affects engineered prompts' scalability across diverse design contexts \cite{fang2024assertllm}. While effective for specific scenarios, one main reason of LLM solutions for hardware security shifting from prompt engineering to fine-tuning (hardware-oriented training) is to better understand and respond the nuances of hardware security queries \cite{fu2023llm4sechw, meng2023unlocking}. However, as fine-tuning typically requires substantial amounts of relevant data to train the model effectively, fine-tuning on hardware, especially due to limited datasets, may present challenges for achieving optimal effectiveness \cite{akyash2024evolutionary}. 

To Address these challenges, this paper introduces Self-HWDebug, a framework that leverages the self-instructional capabilities of LLMs to produce debugging instructions autonomously. In Self-HWDebug, we utilize LLMs to automatically generate debugging instructions by prompting the LLM with pairs of already-crafted vulnerable and secure RTL snippet. These generated instructions are then applied to debug unseen RTL snippets, testing their effectiveness in resolving errors in new and varied hardware configurations but in a same vulnerability categoty.  To mention the core benefits of our approach, we outline the following contributions:

\noindent \underline{(1) Automatic Self-Instructing by LLM:} Self-HWDebug exploits the inherent knowledge embedded within the models, and the automated generation of prompts for better effectiveness, providing instructions with higher specificity/relevance to the tasks compared to expert-crafted instructions.

\noindent \underline{(2) Exploration of References for Self-Instructing:} We examine varying quantities of references used for self-improvement in LLM self-instructing, revealing that the observation of more number of vulnerabilities correlates with an increased success rate in self-instruction for security verification.

\noindent \underline{(3) Scalability and Adaptation:} We evaluate the effectiveness of Self-HWDebug showing the enhanced scalability of the process and allows for rapid adaptation to new vulnerabilities, while circumventing the labor-intensive process of verification.

\section{Background and Related Works}

As with most other research directions, particularly software design and testing, the integration of LLMs is poised to streamline hardware design processes, particularly for electronic design automation (EDA). LLMs, when deployed at higher levels of abstraction, e.g., RTL, offer multifaceted advantages from design to verification: (1) alleviating the burden of manual intervention in implementation tasks \cite{lu2024rtllm, delorenzo2024make}, (2) acting as a substitute for conventional hardware generators, e.g., high-level synthesis (HLS) \cite{liu2024your}, (3) addressing the persistent issue of inadequate HDL codebase availability \cite{lu2024rtllm}, (4) facilitating the acceleration of time-to-market (TTM) in the competitive landscape of IC design \cite{liu2023rtlcoder}, and (5) minimizing the occurrence of human-induced errors \cite{akyash2024evolutionary}.

\begin{figure}[t]
    \centering
    \includegraphics[width=\columnwidth]{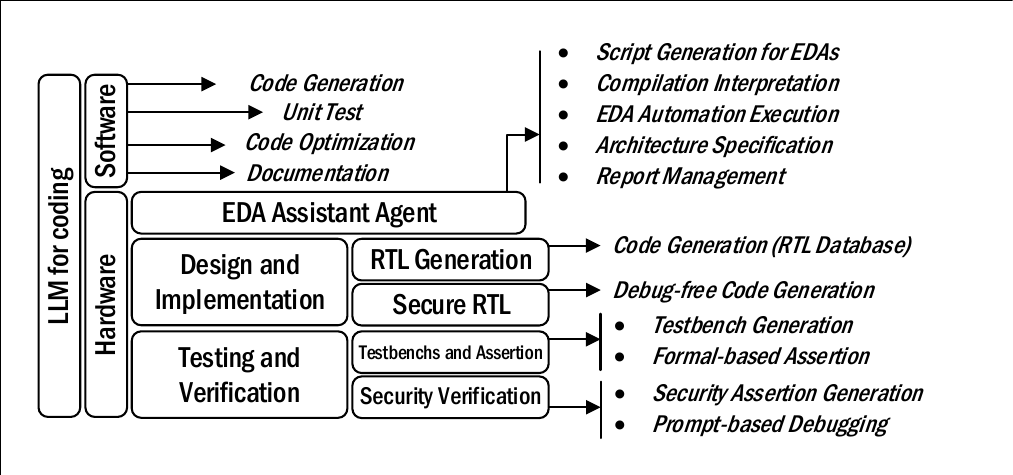}
    \caption{The Use of LLMs for SW/HW Coding (Design) and Test (Verification).}
    \label{fig:llm_coding_tree}
\end{figure}

As shown in Fig. \ref{fig:llm_coding_tree}, Within the domain of hardware, current mechanisms centered around LLMs can be divided into three main groups: (1) Development of automated AI agents tailored to streamline EDA workflows in the IC supply chain, (2) the derivation of software code generation to facilitate RTL implementation, and (3) the use of its semantic parsing for testing and verification. While the first group assists with a range of tasks, e.g., script generation, architecture specification, and interpretation of compilation reports, the second and third group acts as a design and testing assistant to expedite the design and verification process. As shown, a sub-category of the second and third group is centered around the use of LLMs for creating secure RTLs or debugging RTLs with existing vulnerabilities (security-oriented verification). 

\begin{figure}[b]
    \centering
    \includegraphics[width=\columnwidth]{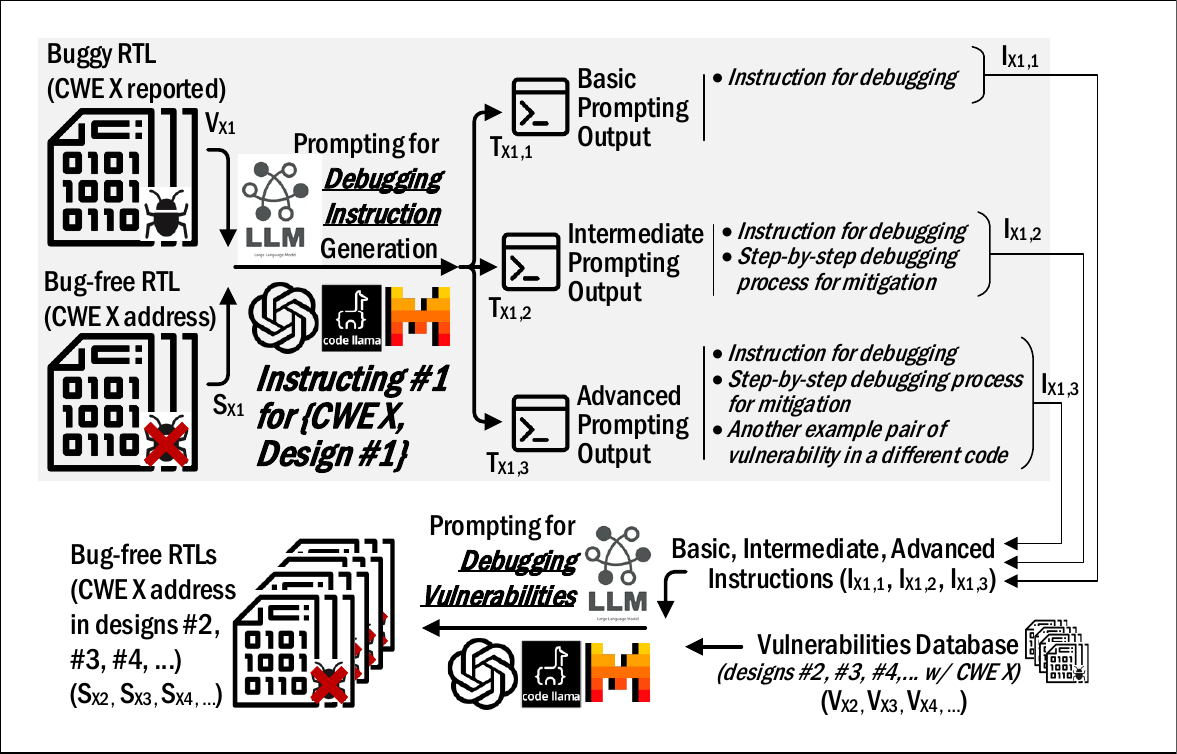}
    \caption{The Overview of Self-Instructing for HW Security Debugging (Based on One-Shot Learning - One Reference for Self-Instructing).}
    \label{fig:llm_self_instruct_1shot}
\end{figure}

\begin{figure}[b]
    \centering
    \includegraphics[width=\columnwidth]{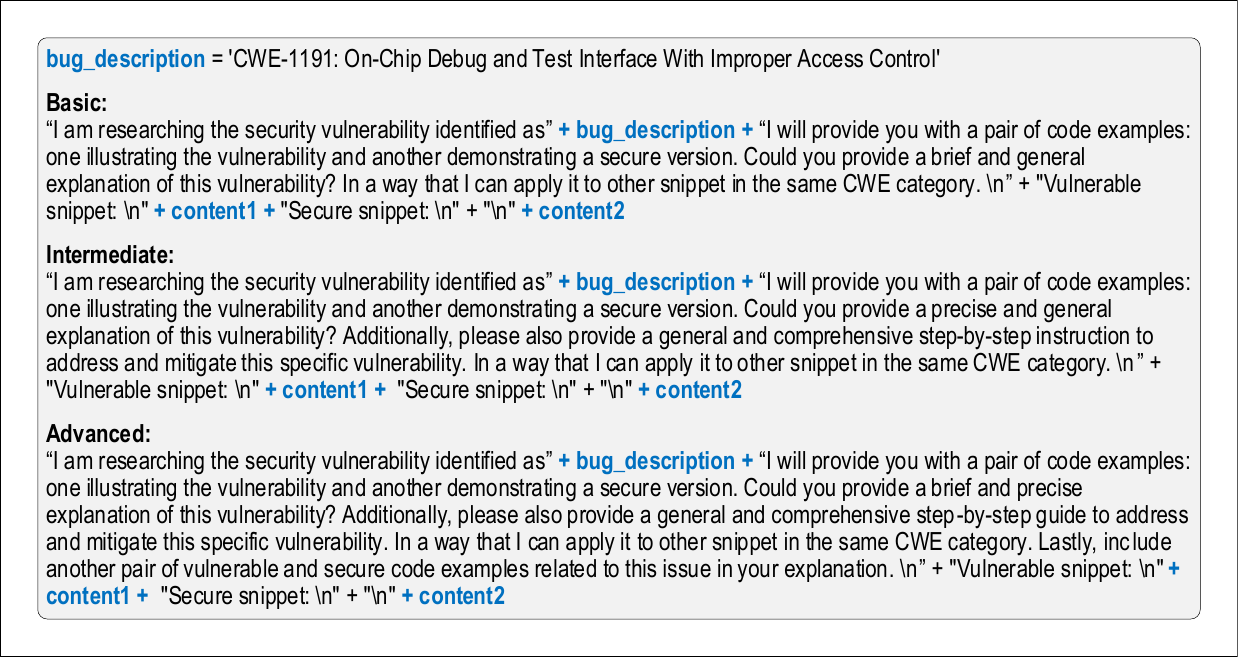}
    \caption{Top View of Task Descriptions at Three Levels (I$_1$-basic, I$_2$-intermediate, I$_3$-advanced) for Instructions' Generation in Self-HWDebug (Sample CWE 1191 for One-shot Learning).}
    \label{fig:llm_inst_levels}
\end{figure}

\begin{figure*}[!b]
    \centering
    \includegraphics[width=\textwidth]{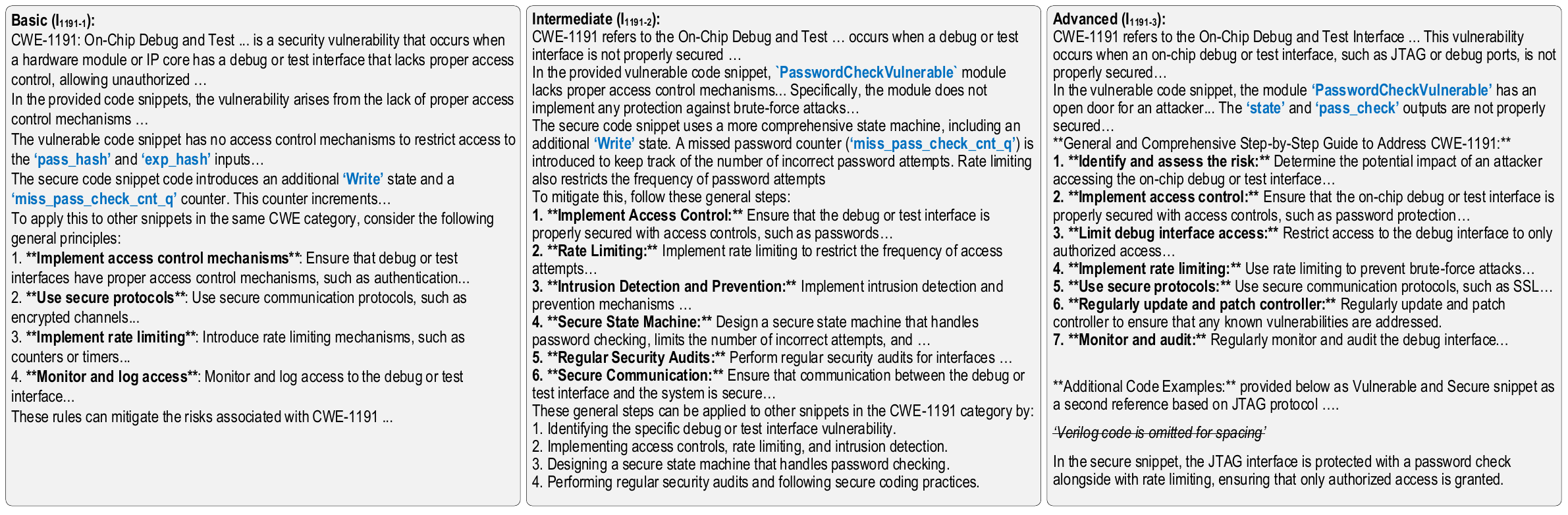}
    \caption{Top View of Generated Instructions (I$_1$-basic, I$_2$-intermediate, I$_3$-advanced) by Llama3 for Sample CWE 1191 based on One-shot Learning.}
    \label{fig:llm_inst_levels_example}
\end{figure*}

In security-oriented solutions, existing studies aim to propel designs towards a state devoid of vulnerabilities, functional or security-related. Analogous to the LLM-based RTL design paradigm, these methodologies fall into two primary streams: (1) prompt engineering, where the refinement of design prompts steers towards generating secure code by LLM \cite{ahmad2024hardware, nair2023generating, orenes2023using}, and (2) RTL-driven tuning, which involves modifying the framework of LLMs themselves based on RTL codebase to yield vulnerability-free outputs \cite{fu2023llm4sechw, meng2023unlocking}.

Although LLMs in hardware security exhibit promising potential, they encounter formidable obstacles in hardware design, testing, and verification. These challenges arise from factors such as prompt engineering and fine-tuning, highlighting the paramount importance of acquiring and effectively utilizing high-quality codebases \cite{gunasekar2023textbooks}. Moreover, the creation of specialized LLMs, e.g., large circuit models (LCMs), or the adaptation of existing models necessitate deep expertise to achieve optimal outcomes in RTL-oriented tasks encompassing generation, detection, and mitigation \cite{li2021prefix, chen2024dawn}. In light of these challenges, the endeavor demands rigorous and concerted efforts across diverse domains.

\section{Proposed Scheme: Self-HWDebug}

In Self-HWDebug, as shown in Fig. \ref{fig:llm_self_instruct_1shot}, generating and testing instructions operates through a two-stage process: First, for a specific vulnerability, e.g., \emph{CWE}$_x$, we generate instructions based on a predefined pair of vulnerable (\emph{V}$_{x1}$) and secure code (\emph{S}$_{x1}$) snippets, known as the reference sample. Using this pair, LLM will be invoked to generate an instruction (\emph{I}$_{x1,i}$), acting as security rule checks, that can be used for debugging the vulnerable snippet. Next, we provide these instructions (\emph{I}$_{x1,i}$) along with a different vulnerable code snippet (\emph{V}$_{xi~|~i>1}$) under the same CWE category (\emph{CWE}$_x$) to the LLM, where the goal is for the LLM to use these generated instructions (\emph{I}$_{x1,i}$) to debug the new vulnerable code and create a secure code snippet (\emph{S}$_{xi~|~i>1}$). Using this approach, the design and verification team can enhance the self-improvement of the LLM and bypass the challenging and time-consuming task of manually crafting instructions by human experts. We utilize one pair (one-shot shown in Fig. \ref{fig:llm_self_instruct_1shot}) and multiple pairs (case study of two-shot shown in Fig. \ref{fig:llm_self_instruct_2shot}) from each vulnerability as reference to produce instructions, while a set of different circuits induced with the same vulnerabilities are used to test the effectiveness of the generated instructions for debugging  RTL codes. 

\subsection{Instruction Generation at Multiple Levels}

Our objective is to develop a set of debugging instructions \emph{I}$_{x1,i}$ for each CWE category, customized to varying levels of detail. These instructions are designed to effectively enable the language model to suggest repairs when paired with instances of vulnerabilities. To do so, for each CWE category (\emph{CWE}$_x$ in Fig. \ref{fig:llm_self_instruct_1shot}), we prepare a task description (\emph{T}$_{x,i}$) along with both vulnerable (\emph{V}$_{x1}$) and secure (\emph{S}$_{x1}$) code snippets. Consider $M$ as the language model. Each instruction is generated in a one-shot manner as $I_{x1,i} = M(T_{x,i} \oplus V_{x1} \oplus S_{x1})$, where $\oplus$ depicts the concatenation of texts\footnote{Note that we use index "1" for \emph{CWE}$_x$ because we use one pair of vulnerable and secure codes for defining the model, i.e., \{\emph{V}$_{x1}$, \emph{S}$_{x1}$\}. The model can be defined with multiple pairs of vulnerable and secure codes, i.e., \{\emph{V}$_{x1}$, \emph{S}$_{x1}$\} and \{\emph{V}$_{x2}$, \emph{S}$_{x2}$\} as shown in Section \ref{sub_sec:multiple_reference} (two-shot).}. For a better comprehensiveness in Self-HWDebug, $i \in \{1, 2, 3\}$ represents the level of detail required from the language model as follows:

\noindent \underline{(1) Basic:} It mostly focuses on a high-level description of the CWE and how it can be basically mitigated.

\noindent \underline{(2) Intermediate:} It covers a high-level of the CWE and how it can be basically mitigated (basic). It also offers a more detailed step-by-step debugging instructions, which resembles a detailed security rule checklist for the targeted vulnerability.

\noindent \underline{(3) Advanced:} It covers a high-level of the CWE and how it can be basically mitigated (basic), alongside with a more detailed step-by-step debugging instructions (intermediate), while it also provides a second example pair of vulnerable and secure code snippet using a different design.

To build a more clear picture of how these levels are generated, and how LLMs are initially invoked for instruction generation at different levels, Fig. \ref{fig:llm_inst_levels} demonstrates a sample showcase on how these levels are defined for task description specialized for CWE-1191. Despite many recent hardware verification techniques centered on LLMs, which demand deep expert knowledge for generation and testing, Self-HWDebug only requires these high-level and generic descriptions, where designers, without needing in-depth security examination and expertise, can generate these descriptions with minimal effort, drawing almost automatically from sources such as CWE databases, design/architecture specification sheets, etc.

\subsection{Mitigating Vulnerabilities with Generated Instructions}

Given the descriptions mentioned above followed by calling the LLM, the required instructions will be generated per vulnerability at the desired levels. Fig. \ref{fig:llm_inst_levels_example} depicts the generated instructions for different levels of detail\footnote{Due to spacing, less critical instruction snippets have been omitted.}. As shown in Fig. \ref{fig:llm_inst_levels_example}, the more advanced instructions provide increased detail regarding the bug and mitigation strategies. These more advanced instructions contribute to a higher success rate in achieving secure design. To see the efficiency of Self-HWDebug, following the generation of these instructions, they are utilized to test unseen vulnerable code snippets within the same vulnerability category, aiming to mitigate the vulnerability across various designs. Assume we possess code ($V_{xi}$) that contains a security vulnerability \emph{CWE}$_x$. We obtain the secure snippet ($S_{xi}$) by prompting the LLM with the corresponding instruction and the vulnerable code as $S_{xi} = M(T_g \oplus I_{xi} \oplus V_{x})$. Where $T_g$ is a general task description that we ask LLM to mitigate the vulnerability according to the given instruction. 
    
\subsection{Using multiple references for higher accuracy}
\label{sub_sec:multiple_reference}

When employing a single reference (one-shot) to generate instructions (e.g., \emph{V}$_{x1}$ and \emph{S}$_{x1}$ for \emph{CWE}$_x$), our experimental results show acceptable success rate. However, depending on the type of the vulnerability, difficulties to mitigate, vulnerability and task/instruction representation, etc., different vulnerabilities within the same CWE category can require distinct mitigation techniques, and one technique (one reference) may not sufficiently address the same vulnerability in a different design. Recognizing this limitation in the one-shot method (Fig. \ref{fig:llm_self_instruct_1shot}), where the instruction might not include multiple techniques, we explore the possibility of using multiple references (showcasing two-shot), targeting a higher success rate (better coverage of the same vulnerability while the LLM targets the same CWE across a wider array of unseen designs). 

\begin{figure}[t]
    \centering
    \includegraphics[width=\columnwidth]{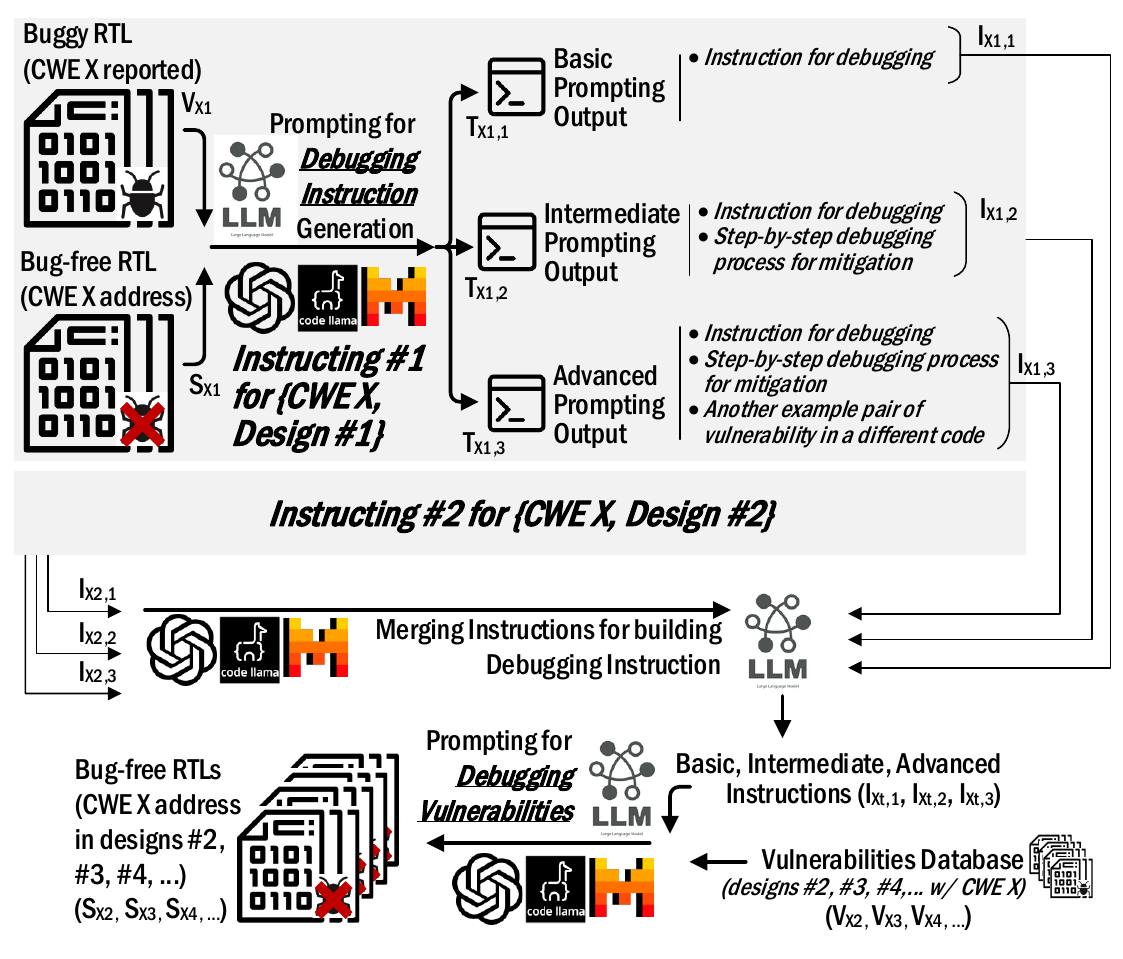}
    \caption{The Overview of Self-Instructing for HW Security Debugging (Based on Two-Shot Learning - Two References for Self-Instructing).}
    \label{fig:llm_self_instruct_2shot}
\end{figure}

To enable the use of multiple references (e.g., two-shot), as shown in Fig. \ref{fig:llm_self_instruct_2shot}, we prompt the LLM to generate instructions using two distinct pairs of references that address the same vulnerability through varied techniques/designs (\emph{CWE}$_x$ in both \{\emph{V}$_{x1}$, \emph{S}$_{x1}$\} and \{\emph{V}$_{x2}$, \emph{S}$_{x2}$\}). To obtain the two-shot instruction (\emph{I}$_{xt}$ as the combination of \emph{I}$_{x1}$ and \emph{I}$_{x2}$), we prompt the LLM with a task description (see Fig. \ref{fig:llm_self_instruct_2shot_example}) formulated as \(I_{xt} = M(T_{xt} \oplus V_{x1} \oplus S_{x1} \oplus V_{x2} \oplus S_{x2})\), where (\emph{V}$_{x1}$, \emph{S}$_{x1}$) and (\emph{V}$_{x2}$, \emph{S}$_{x2}$) are our reference samples (two pairs of vulnerable and secure snippet codes). By contrasting the task descriptions from the one-shot (Fig. \ref{fig:llm_inst_levels}) and two-shot approaches (Fig. \ref{fig:llm_self_instruct_2shot_example}), the two-shot method prompts the LLM to generate a \textit{combined} instruction that considers both references (to build a more comprehensive set of instructions, i.e., \emph{I}$_{xt,i=1,2,3}$. In our experimental results, by analyzing The instructions generated using in two-shot approach, we demonstrate that using multiple reference (here is 2), the mitigation operates more comprehensive and contain more detail and strategies compared to their one-shot counterpart. It is noteworthy that an increasing number of references could potentially enhance comprehensiveness. However, in future studies, we demonstrate that there is a threshold for the number of references beyond which the model may be misled.

\section{Experiments and Results}

To assess the effectiveness of Self-HWDebug, we conduct experiments that involve generating and testing instructions in both one-shot and two-shot formats over a set of CWEs. Furthermore, we explore various levels of instruction complexity to gauge the impact of more advanced directives. Additionally, we experiment with integrating guidance from an expert LLM to determine whether their advanced knowledge can improve the mitigation efforts of an open-source model.

\subsection{Bugs Descriptions}

\begin{figure}[t]
    \centering
    \includegraphics[width=\columnwidth]{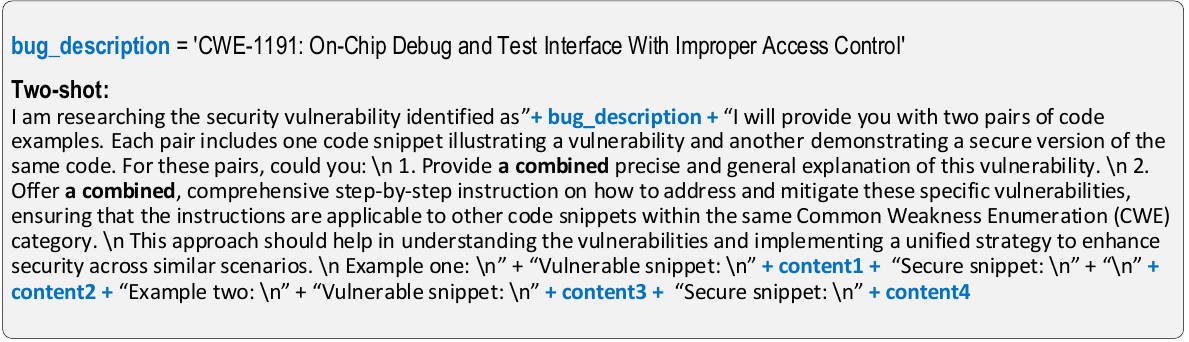}
    \caption{Generating Instruction for Two-shot Learning (Sample CWE 1191).}
    \label{fig:llm_self_instruct_2shot_example}
\end{figure}

In our experimental setup, we focused on a subset (five) of most important hardware CWEs \cite{mitre_cwe_bugs}. For each category, a database of seven distinct samples (snippet code with vulnerabilities) based on the respective CWE descriptions (from different sources, e.g., MITRE \cite{mitre_cwe_bugs}, hackathons \cite{sadeghi2021organizing}, Trust-hub \cite{trust-hub}, etc.). We designated one/two of these samples as the reference (shot) for generating instructions, while we used five (different designs) to test the efficacy of the generated instructions in addressing and rectifying the issues. Table \ref{tab:bug_description} demonstrates the utilized CWE categories and their description. For two-shot approach, we used and extra pair of vulnerable and secure code as the reference.

\subsection{Experimental Settings}

In all our experiments we utilized recently-introduced Llama3-70B \cite{llama3_2024}. This is an open-source high-capacity language model from Meta, designed for text and code generation. For implementing LLM, we employ the Groq API \cite{groq2024}. This API uses Groq's cutting-edge LPU technology, which provides extremely fast AI inference capabilities, and make it highly suitable for tasks that require real-time performance. As of May 2024, Groq provides a free plan, though it comes with some limitations. To ensure consistency in our analysis of the experiments, we set the temperature and top-p parameters to constant values of 0.6 and 1, respectively. This approach allows us to examine the effects of other variables without the influence of the probabilistic nature of LLM outputs.

\subsection{Instruction Generation in One- and Two-Shot Approaches}

In the one-shot approach, we prompt the LLM with a single reference to generate general debugging instructions at varying levels of detail\footnote{To enhance the accuracy of these instructions, a hardware designer can add general annotation to the reference code with comments.}. Our observations indicate that the LLM-produced instructions are comprehensive, capturing both the essence of the vulnerability and the necessary debugging steps for mitigating the vulnerability. These instructions are then validated: the LLM is prompted with an unseen snippet to provide a repair solution, and we use an assertion-based validation on the the repaired code to verify its validity.


\begin{table}[t]
\fontsize{7pt}{8pt}\selectfont
\centering
\caption{Description of the utilized CWE categories.}
\label{tab:bug_description}
\setlength\tabcolsep{2pt} 
\begin{tabular}{@{} p{50pt} p{195pt} @{}}
\toprule  
CWE category & Bug description  \\
\cmidrule(r){1-1}\cmidrule(r){2-2}
CWE-1191 & Pertains to vulnerabilities in on-chip debug and test interfaces that lack proper access controls, potentially allowing unauthorized access or manipulation of the chip's functions. \\
\cmidrule(r){1-1}\cmidrule(r){2-2}
CWE-1231 & Refers to vulnerabilities arising from the improper prevention of modifications to lock bits, which can lead to unauthorized changes in the device's functionality or security settings. \\
\cmidrule(r){1-1}\cmidrule(r){2-2}
CWE-1244 & Involves vulnerabilities where internal assets are exposed due to being set at an unsafe debug access level or state, potentially compromising the security and integrity of the system. \\
\cmidrule(r){1-1}\cmidrule(r){2-2}
CWE-1245 & Refers to vulnerabilities due to improperly designed Finite State Machines (FSMs) in hardware logic, which can lead to unpredictable behavior or security risks in the hardware's operation. \\
\cmidrule(r){1-1}\cmidrule(r){2-2}
CWE-1300 & Relates to vulnerabilities from inadequate safeguards against physical side channels, which can inadvertently reveal critical information through the hardware's electromagnetic signals, acoustic outputs, or power consumption patterns. \\
\bottomrule
\end{tabular}
\end{table}

\begin{figure*}[t]
    \centering
    \includegraphics[width=\textwidth]{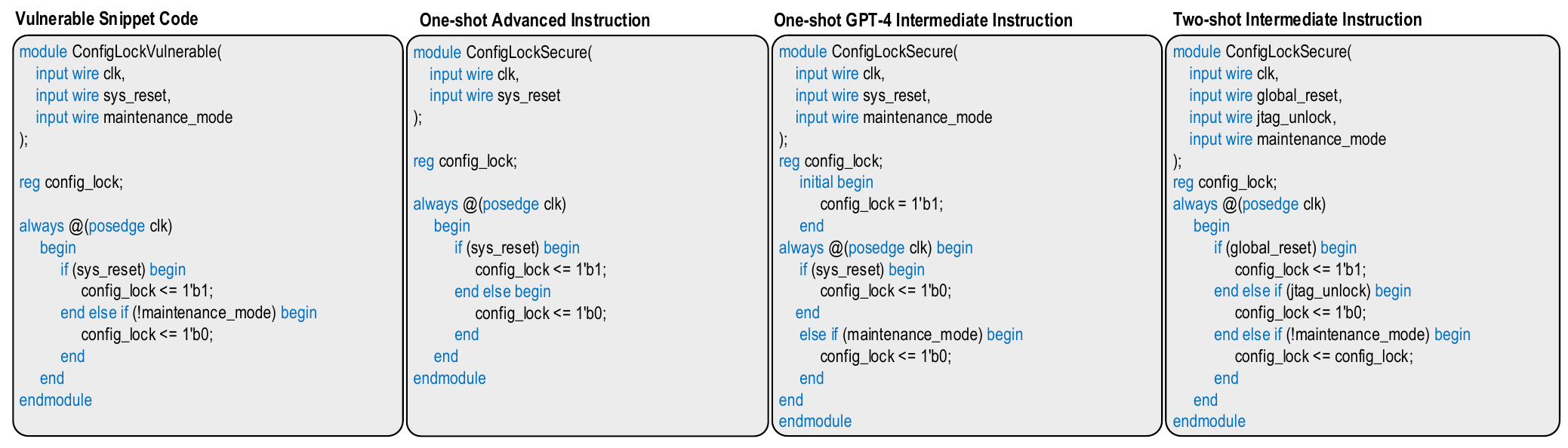}
    \caption{One example of a vulnerable code snippet (CWE 1231) alongside its repairs, each generated with different levels of instruction detail.}
    \vspace{-15pt}
    \label{fig:bug_levels_example}
\end{figure*}

\subsection{Instruction Generation by More Advanced Model}

Apart from the instruction and testing generation all done by LLama3, we also incorporated knowledge from more advanced and capable models, e.g., GPT-4, to determine if it can further improve mitigation success rate. For this part of the experiment, we utilized GPT-4 to generate detailed instructions based on a reference pair and then apply these instructions to Llama3 for generating the repairs. 

Our findings, as shown in Table \ref{tab:debug_accuracy}, demonstrate that GPT-4's comprehensive instructions greatly aid Llama3 in performing more effective code repairs. This approach allows us to leverage the advanced capabilities of a limited-access, closed-source model (i.e., GPT-4) to augment an open-source model (i.e., Llama3) and achieve comparable performance at a lower cost. The integration of GPT-4 involves a process similar to knowledge distillation, where GPT-4, serving as a sophisticated \emph{'teacher'}, transfers complex debugging strategies and subtle details to Llama3, the \emph{'student'}. This method infuses Llama3 with enhanced capabilities to handle complex debugging tasks that were previously out of reach and demonstrate a practical application of knowledge distillation in bridging the gap between proprietary and open-source LLMs.

\subsection{Comparison of Different Levels of Instruction}

Table \ref{tab:debug_accuracy} depicts the mitigation results for the targeted CWE categories and different instructions' levels. As demonstrated, the model exhibits improvement from basic to advanced task descriptions in one-shot self-instructing. However, this improvement is not always consistent, as the quality of the model's responses can vary due to inherent randomness and certain limitations. Nonetheless, by transitioning to a more advanced model, such as GPT-4, and employing a two-shot model approach (moving towards optimum -multiple- number of references), even at an intermediate level of task description, the success rate remains consistently high.

To draw a top picture of how this self-instructing performs for a specific scenario, Fig. \ref{fig:bug_levels_example} shows a snippet code of CWE 1231 and its mitigation approaches with different level of details (based on the instructions generated from Figs. \ref{fig:llm_inst_levels} and Fig. \ref{fig:llm_self_instruct_2shot_example}). In the vulnerable snippet code the logic mistakenly sets the \texttt{config\_lock} to \texttt{1'b0} (unlocked) when the state is not in \texttt{maintenance\_mode}, which is counter-intuitive as it should remain locked to protect the system configuration. This leaves the system vulnerable to unauthorized changes almost at normal (functional) mode (vs. during the test mode). 

In the advanced level mitigation, LLM simplifies the control logic by removing the maintenance mode check, resulting in a configuration that is always unlocked except during the reset. This approach is less secure than even the flawed original, as it does not attempt to verify the context or condition under which unlocking is permissible. With instructions generated by GPT-4, the LLM tries to mitigate the bug by unlocking during both system resets and maintenance mode which enables a more flexible and accessible system management. With the two-shot approach, the LLM (all based on Llama3) introduces a multi-condition lock control mechanism, leveraging both a global reset and a specific JTAG unlock condition, which can be tied to authenticated sessions or cryptographic checks.

\subsection{Takeaways for Self Instructing in Hardware Verification}

\begin{table}[t]
\fontsize{6.5pt}{8pt}\selectfont
\centering
\begin{threeparttable}
\caption{Efficacy Ratio of Self-HWDebug in Self-Instructing for Security Verification (Debugging using One/Two-Shot Learning).}
\label{tab:debug_accuracy}
\setlength\tabcolsep{2pt} 
\begin{tabular}{@{} p{40pt} p{35pt} p{40pt} p{35pt} p{35pt} p{40pt}@{}}
\toprule  
Vulnerability & Basic & Intermediate & Advanced  & GPT-4$^{*1}$ & Two-shot$^{*1}$ \\
\cmidrule(r){1-1}\cmidrule(r){2-2}\cmidrule(r){3-3}\cmidrule(r){4-4}\cmidrule(r){5-5}\cmidrule(r){6-6}
CWE-1191 & 2 out of 5 & 5 out of 5 & 3 out of 5 & 4 out of 5 & 5 out of 5\\
\cmidrule(r){1-1}\cmidrule(r){2-2}\cmidrule(r){3-3}\cmidrule(r){4-4}\cmidrule(r){5-5}\cmidrule(r){6-6}
CWE-1231 & 0 out of 5 & 0 out of 5 & 1 out of 5 & 2 out of 5 & 5 out of 5 \\
\cmidrule(r){1-1}\cmidrule(r){2-2}\cmidrule(r){3-3}\cmidrule(r){4-4}\cmidrule(r){5-5}\cmidrule(r){6-6}
CWE-1244 & 5 out of 5 & 4 out of 5 & 5 out of 5 & 5 out of 5 & 5 out of 5 \\
\cmidrule(r){1-1}\cmidrule(r){2-2}\cmidrule(r){3-3}\cmidrule(r){4-4}\cmidrule(r){5-5}\cmidrule(r){6-6}
CWE-1245 & 5 out of 5 & 5 out of 5 & 5 out of 5 & 5 out of 5 & 5 out of 5\\
\cmidrule(r){1-1}\cmidrule(r){2-2}\cmidrule(r){3-3}\cmidrule(r){4-4}\cmidrule(r){5-5}\cmidrule(r){6-6}
CWE-1300 & 2 out of 5 & 4 out of 5 & 5 out of 5 & 5 out of 5 & 5 out of 5 \\
\cmidrule(r){1-1}\cmidrule(r){2-2}\cmidrule(r){3-3}\cmidrule(r){4-4}\cmidrule(r){5-5}\cmidrule(r){6-6}
Average & 56\% & 72\% & 76\% & 84\% & 100\% \\
\bottomrule
\end{tabular}
\begin{tablenotes}
\item[$^{*1}$] For both GPT-4 and two-shot experiments, we used an intermediate level of detail when prompting the LLM.
\end{tablenotes}
\end{threeparttable}
\end{table}

\textit{Dependence of Mitigation on the Accuracy of the Reference Sample:} Instructions and subsequent mitigation become most effective when the reference examples accurately represent the issue across various scenarios. Therefore, using multiple reference pairs, each employing different approaches for mitigation, leads to more comprehensive and general instructions and thus, more effective self-instruction-based mitigation.

\textit{Detailed Instructions Lead to More Sophisticated Repairs:} As the complexity and detail in the instructions or descriptions increase, particularly while coupled with another example within the instruction, the potential for innovative and sophisticated repairs also rises. This comprehensive understanding allows LLM to devise more complex and effective solutions.

\textit{Variability in Mitigation Difficulty Based on the Vulnerability:} The difficulty of mitigating a vulnerability can vary significantly depending on the nature of the vulnerability itself. Some vulnerabilities might be straightforward , while others might be inherently complex to mitigate. Depending on their nature, multiple/advanced instructions are required to guarantee a high success rate in Self-HWDebug. 

\section{Conclusion and future work}

In this paper, we introduce a new framework designed to enhance the scalability and efficiency of LLMs in mitigating security vulnerabilities in hardware designs in a more automated manner. Prior research has shown that LLMs can effectively manage vulnerability mitigation. However, generating detailed, hand-written instructions has remained a significant challenge. This is primarily because the creation of these instructions demands significant time and effort from experts and typically only reflects the knowledge scope of a hardware engineer. To address these limitations, our proposed solution, Self-HWDebug, leverages the LLM's ability to autonomously generate instructions. This approach not only addresses scalability issues but also produces instructions that are comprehensive and informed by an expanded knowledge base. In Self-HWDebug, the LLM acts as its own teacher and facilitates self-improvement. Through an initial testing, our experiments shows the efficacy of Self-HWDebug over a sub-set of CWE vulnerabilities with high success rate.

We view Self-HWDebug as a significant advancement in automating the mitigation of security vulnerabilities using LLMs, though require further study. Firstly, we plan to evaluate it over a wide (complete) range of CWEs and expand our dataset to advance the LLM's capabilities towards more comprehensiveness. Also, we are considering the possibility of using the LLM both as a detector and mitigator of bugs. We believe this approach is feasible and would allow the application of LLMs to large-scale SoC designs, rather than being limited to snippets of already-detected vulnerabilities.

\bibliographystyle{IEEEtran}
\bibliography{refs}

\end{document}